\documentstyle[multicol,aps,epsfig]{revtex}

\newcommand{\be}{\begin{equation}}
\newcommand{\ee}{\end{equation}}
\newcommand{\bea}{\begin{eqnarray}}
\newcommand{\eea}{\end{eqnarray}}

\newcommand{\delsq}{\nabla^2}

\newcommand{\fd}[2]{\frac{\partial #1}{\partial #2}}

\newcommand{\ket}[1]{| #1 \rangle}

\begin{document}
\draft

\title{Self-consistent model of ultracold atomic collisions and 
Feshbach resonances in tight  harmonic traps}
\author{E. L. Bolda, E. Tiesinga, and P. S. Julienne}
\address{Atomic Physics Division, 
National Institute of Standards and Technology,
100 Bureau Drive, Stop 8423, 
Gaithersburg MD 20899-8423} 
\date{January 10, 2002}

\maketitle

\begin{abstract}
We consider the problem of cold atomic collisions in tight traps,
where the absolute scattering length may be larger than the trap size. 
As long as the size of the trap ground state is larger than a
characteristic length of the van der Waals potential, the energy
eigenvalues can be computed self-consistently from the scattering
amplitude for untrapped atoms.  By comparing with the exact numerical
eigenvalues of the trapping plus interatomic potentials, we verify
that our model gives accurate eigenvalues up to milliKelvin energies
for single channel $s$-wave scattering of $^{23}$Na atoms in an
isotropic harmonic trap, even when outside the Wigner threshold
regime.  Our model works also for multi-channel scattering, where the
scattering length can be made large due to a magnetically tunable
Feshbach resonance.

\end{abstract}

\pacs{PACS numbers: 32.80.Pj, 32.80.Lg, 34.50.-s, 39.25.+k}

\begin{multicols}{2}
\narrowtext
    
\section{introduction}

Along with the development of laser cooling of atoms have come
techniques for trapping the cold atoms, with tremendous advantages for
experimental atomic physics.  Just to name a few potent examples where
trapping is necessary, the Bose-Einstein transition has been reached
in several atomic species  \cite{Dalfovo99,Inguscio99,Parkins98}, 
threshold scattering properties have been studied 
\cite{Julienne01k,Weiner99},  molecules formed with the assistance of 
light \cite{Stwalley99}, and quantum chaos 
\cite{Raizen00,Hensinger01} 
and quantum phase transitions \cite{Greiner02} observed using optical 
lattices.  
a 




Atomic collisions play an essential role in most of these phenomena.  
In the past one could ignore the fact that these collisions take 
place in a 
trap, since trap sizes are very large in comparison with the sizes 
associated with atomic interactions.  However, recent developments 
make it crucial to account for the effect of trap confinement on 
collisions when the atoms are held tightly in one, two, or three 
dimensions by optical lattices.  For example, Greiner et al.\ 
\cite{Greiner02} have observed a quantum phase transition from a 
superfluid to 
a Mott insulator within a 
three-dimensional optical lattice. Moreover, several 
low-dimensional phase transitions of cold bosonic systems have been 
conjectured.  In two dimensions, the
Kosterlitz-Thouless transition may occur \cite{Stoof93}, while in one
dimension the Tonks-Girardeau phase should be possible
\cite{Olshanii98,Dunjko01}.  Zero-temperature transitions have
also been investigated theoretically for a rotating two dimensional
gas \cite{Mottelson99}.  All of these transitions depend on atomic 
collisions, and for
quantitative predictions the low-dimensional interactions must 
be understood. 

Two proposals for quantum computing involve loading cold atoms into
optical lattices, and using the interaction between the atoms as the
switching mechanism \cite{Brennen99,Jaksch99}. 
In one type of quantum logic gate, two atoms are brought together and 
allowed to
interact for a set time interval, resulting in different phase shifts
depending on their hyperfine sublevels.  A recent experiment 
\cite{Greiner02} represents an important first step towards quantum 
logic applications, since it shows that a lattice can be initialized 
with uniform occupancy of lattice sites.

Another burgeoning area is the study of Feshbach resonances, and
weakly-bound molecular states, in the interaction of two ultra-cold 
atoms.  This has improved
the knowledge of interaction parameters of alkali atoms and opened up
the field of molecular condensates and 3-body processes
\cite{Feshbachexpt,Timmermans99a}.  By tuning Feshbach
resonances one can easily reach an interesting regime where the
scale length associated with the scattered wave exceeds the trap width
\cite{Tiesinga00a}.

We address these problems by calculating the eigenvalues of two 
interacting atoms confined in a trapping potential.  A popular method 
for representing cold atom interactions is to replace the exact 
interatomic potential by a delta-function pseudopotential 
proportional to the scattering length $a$:
\be
  {\hat V} = 
  \frac{4 \pi \hbar^2}{m} a \delta({\bf r}) 
  \fd{}{r} r,
  \label{Eq1}
\ee
where $m$ is the atomic mass and $r$ is the interatomic separation.
An analytic solution for the eigenvalues of two atoms in an isotropic 
harmonic trap plus the pseudopotential Eq.~(\ref{Eq1}) has been found 
\cite{Busch98a}.  However, some of us have
previously shown that the use of this solution is limited to
sufficiently weak traps such that the trap width is much larger
than $|a|$ \cite{Tiesinga00a}.   Here we reexamine this
problem and propose a self-consistent method of calculating the trap
energies, which gives good quantitative results over a wide range of
trap frequencies, even when $|a|$ is larger than the trap size.  The 
essence of our model is to replace $a$ with an energy-dependent 
effective scattering length.   An advantage of our model  is that 
once the 
energy-dependent scattering phase shift
for a particular type of cold collision is known, either from
experiment or from close-coupling calculations, it can be easily
applied to obtain eigenvalues for traps of all frequencies.
Conversely, if the eigenvalues are measured, information about
collisions can be obtained. 

We note that the pseudopotential can be used to obtain approximate
solutions for trapped colliding atoms in one dimension
\cite{Petrov00b,Olshanii98} and two dimensions \cite{Petrov00a}.  It 
may be possible to adapt our self-consistent method to accurately 
treat scattering in ``cigar-'' or ``pancake-'' shaped traps.

The paper is organized as follows.  In Sec.  II we formulate the
problem of atoms colliding in a tight spherical trap, and briefly
review scattering theory.  In Sec.  III we motivate and explain the
self-consistent eigenvalue model, which is our main result. 
Limitations of the model are discussed.  Section  IV applies the 
model to single-channel scattering of $^{23}$Na atoms in a trap, and 
shows good agreement  with numerical calculations using the full
interaction Hamiltonian.  Section V demonstrates similar good 
agreement for the case of multi-channel scattering.  Specifically, we 
consider a 
magnetically-induced Feshbach resonance in Na$_2$.  We compare, for a 
range of magnetic  fields, exact numerical results from the  
five-channel close-coupled  scattering problem with the 
self-consistent eigenvalues.  Finally, in  Sec.  VI we draw 
conclusions and consider more general traps and  applications to 
many-body theory.

\section{Two atoms colliding in an isotropic harmonic trap}

We consider an isotropic harmonic trap described for atom $j=1,2$ at 
position 
${\bf r}_j$ by 
\be
V_{\mbox{trap}}({\bf r}_j) = \frac{1}{2} m \omega^2 r_j^2,
\ee
where $\omega$ is the trapping frequency.  Harmonic traps can be made 
by a variety of means.  Very tight confinement is possible with a 
three-dimensional optical lattice.  Typical experimental trap 
frequencies range from $50$ kHz to $1$ MHz. These optical dipole 
traps are much tighter than those obtained with magnetic fields. 
In a recent experiment \cite{Greiner02}, isotropic potentials at each 
site were produced from three optical standing waves of equal 
intensity.

For the isotropic harmonic trap, the two-atom Hamiltonian is 
separable in
the center-of-mass and relative coordinates.  Since the
center-of-mass motion is just that of the well-known isotropic 
harmonic oscillator, we need only discuss the problem in the 
relative coordinates.  The Hamiltonian is
\be
H = - \frac{\hbar^2}{2 \mu} \delsq + \frac{1}{2} \mu \omega^2 r^2 + 
V_{\mbox{int}}(r),
\label{Hamiltonian}
\ee
where $r = | {\bf r}_1 - {\bf r}_2 |$, $\mu = m/2$ is the reduced 
mass, and $V_{\mbox{int}}(r)$ is the interatomic potential. In 
(relative) spherical
coordinates, the trap states neglecting $V_{\mbox{int}}(r)$ have
energy eigenvalues
\be
E_n^{(0)} = \left( 2 n + L + \frac{3}{2} \right) \hbar \omega,
\label{unperturbE}
\ee
where $n = 0,1,2, \ldots$ is the radial quantum number and $L=0,1,2, 
\ldots$ is the partial wave quantum number.   We henceforth consider 
only $s$-waves ($L=0$).
The size of the ground state trap wavefunction is characterized by 
\be
l = \sqrt{\frac{\hbar}{\mu \omega}}.
\ee
Typical trap sizes $l$ for Na in the above mentioned trap 
frequency range are $30$ nm to $130$ nm.  

The interatomic potential $V_{\mbox{int}}(r)$  is characterized by a 
short-range region of strong chemical bonding and a long-range van 
der Waals potential,
\be
V_{\mbox{int}} \rightarrow -C_6/r^6,
\ee
and leads to a van der Waals scale length, 
\cite{Weiner99,Gribakin93,Williams99}
\be
 x_{0} = \frac{1}{2} \left(\frac{2\mu C_6}{\hbar^2}\right)^{1/4}.
\ee
For $r \ll x_0$ the scattering wavefunction oscillates rapidly due to 
the strong interaction potential.  In alkali ground state 
interactions, $C_6$ is the same for all hyperfine 
states of a given atomic pair; consequently, $x_0$ is the same for 
all collision channels.  In the case of Na$_2$ considered below, it 
is about 2.4 nm. 

For collisions of atoms in the absence of a trapping potential, the 
asymptotic $s$-wave scattering wavefunction for relative collision 
momentum $\hbar k$ approaches
\be
\psi \rightarrow  \frac{\sin(kr + \delta_0)}{\sqrt{k}r} 
\label{asympwave}
\ee
at large interatomic separation $r \gg x_0$.  Another length scale 
that
naturally appears for cold collisions is the scattering length, 
defined in terms of the $s$-wave phase shift $\delta_0$ by
\be 
a = - \lim_{k \rightarrow 0} \frac{\tan \delta_0(k)}{k}.
\label{ascat}
\ee
The Wigner law regime is then defined by the range of momenta for 
which $\delta_0 =  ka$ is a good approximation, i. e.,
\be
k  \ll \frac{\pi}{2|a| }.
\label{Wignerregime}
\ee
The scattering length can take on any value between $+\infty$ and 
$-\infty$.    As $|a|$ becomes large, the range of $k$ for which the 
Wigner law applies becomes very small.

In view of typical trap sizes and van der Waals length scales, we need
only consider the experimentally accessible regime, for which
\be
x_{0} \ll l.
\label{xGFlessthanl}
\ee
On the other hand, the scattering length can
have a larger magnitude than the trap width $l$.  This is especially
likely if the scattering length is modified by means of a Feshbach
resonance. 

Our goal is to find a simple model for calculating the new energy
eigenvalues of the trap when collisions are present.  An analytical
solution of this problem was presented in Ref.  \cite{Busch98a} by 
replacing $V_{\mbox{int}}$ by the pseudopotential of 
Eq.~(\ref{Eq1}).  This replacement assumes that the Wigner law is 
valid.  However, we previously showed that the eigenvalues thus 
obtained are not always in agreement with numerical results 
\cite{Tiesinga00a}!  Specifically, they are least accurate when $|a|$ 
approaches or exceeds $l$.  One way to see this is that the energies 
of the unperturbed trap states are already large enough that the 
Wigner threshold law is invalid.  For the unperturbed trap ground 
state $E =3 \hbar \omega /2$ and hence the root-mean-square momentum 
$k = \sqrt{3}/l$.  Therefore by Eq.~(\ref{Wignerregime}) we are 
outside of the Wigner regime if $|a| > \pi/(2 \sqrt{3})\,l$.

In the next section we will use the inequality 
Eq.~(\ref{xGFlessthanl})
to motivate a self-consistent model of cold collisions in the trap,
that is valid at all relevant energies and scattering lengths.

\section{self-consistent model}

The improved model we propose relies on a generalization of the
pseudopotential approximation for $V_{\mbox{int}}$ in 
Eq.~(\ref{Eq1}). 
We introduce the energy-dependent pseudopotential operator
\cite{Huang57}
\be {\hat V}_{\mbox {eff}}  = 
  \frac{4 \pi \hbar^2}{m} a_{\mbox{eff}}(E) \delta({\bf r}) 
  \fd{}{r} r,
  \label{pseudopotential}
\ee
where the {\em effective} scattering length is defined as
\be
a_{\mbox {eff}}(E)  = - \frac{\tan \delta_0(k)}{k}.
\label{aeff}
\ee
and the kinetic energy is related to the momentum by $E = \hbar^2 
k^2/2 \mu$. 
This operator gives the same asymptotic wavefunction, 
Eq.~(\ref{asympwave}), as the full interaction potential 
$V_{\mbox{int}}$. 
The effective scattering length reduces to the usual one,
Eq.~(\ref{ascat}), in the Wigner threshold regime.  The phase shift in
Eq.~(\ref{aeff}) does not need to be small in order to use 
Eq.~(\ref{pseudopotential}).  Even though the effective scattering 
length
diverges when $\delta_0$ is an odd multiple of $ \pi/2$, the
wavefunction remains well-behaved.

Reference~\cite{Busch98a} found the eigenvalues of the trapped atoms 
interacting
through the operator in Eq.~(\ref{Eq1}) as the solutions of the 
equation,
\be
\frac{a}{l} = f(E) ,
\label{analyticeigen}
\ee
where the ``intercept function''  is
\be
f(E) = \frac{1}{2} \tan \left(\frac{\pi E}{2 \hbar \omega} + 
\frac{\pi}{4} \right) \frac{ \Gamma \left(\frac{E}{2 \hbar \omega} + 
\frac{1}{4} \right)}{      \Gamma \left(\frac{E}{2 \hbar \omega} + 
\frac{3}{4} \right)}  
\label{interceptfn}
\ee 
and $\Gamma$ is the gamma function.  To account properly for the
scattering in tight traps, where the Wigner law may not apply at the 
trap energies, we need to replace Eq.~(\ref{analyticeigen}) by one in 
which the left-hand side is energy-dependent and 
solve the  equation  
\be
\frac{a_{\mbox {eff}}(E)}{l} = f(E)
\label{selfconsEq}
\ee
self-consistently for the eigenvalues.

One might ask, why does the idea of the pseudopotential still work
outside the regime of the Wigner law?  The answer is that the
collision occurs on the very short length scale $x_0$, so the
interatomic interaction potential is undistorted by the trap.  This in
turn means that the kinetic energy at which the effective scattering
length needs to be evaluated is the eigenvalue itself, since the trap
potential is negligible for $r < x_0$.  Thus we were led to the
self-consistent Eq.~(\ref{selfconsEq}).

This model can be expected to break down if the trap becomes too
tight.  The interatomic potential $V_{\mbox{int}}$ becomes comparable
to the trap potential near $r =\sqrt{l x_{0}}$.  Hence the inner part
of the wavefunction where the scattering occurs is nearly the same as
that without the trap when $x_{0} \alt \sqrt{l x_{0}}$, equivalent to
Eq.~(\ref{xGFlessthanl}).  A different kind of limitation is that
this model cannot predict bound states without our knowing the
analytical continuation of the effective scattering length to negative
energies.

\section{single-channel scattering}
\label{singlesect}
\begin{figure}[tb]
\noindent\centerline{ 
\epsfig{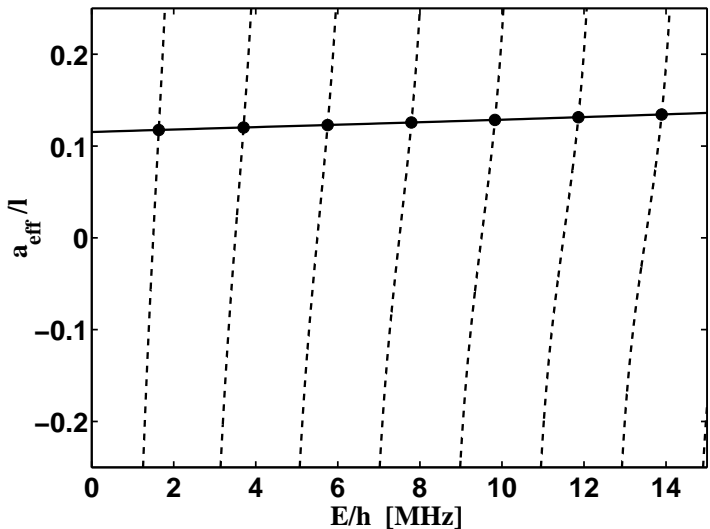} }
    \caption{Effective scattering length (solid curve) and intercept
    function $f(E)$ (dashed) versus energy for doubly polarized 
    $^{23}$Na in a 1
    MHz trap.  The energies at which the two curves intersect give the
    self-consistent eigenvalues.  The circles show the actual 
positions of
    the exact numerical eigenvalues along the curve of the intercept
    function.  }
 \label{selfconsplot1}
 \end{figure}

\begin{figure}[tb]
\noindent\centerline{ 
\epsfig{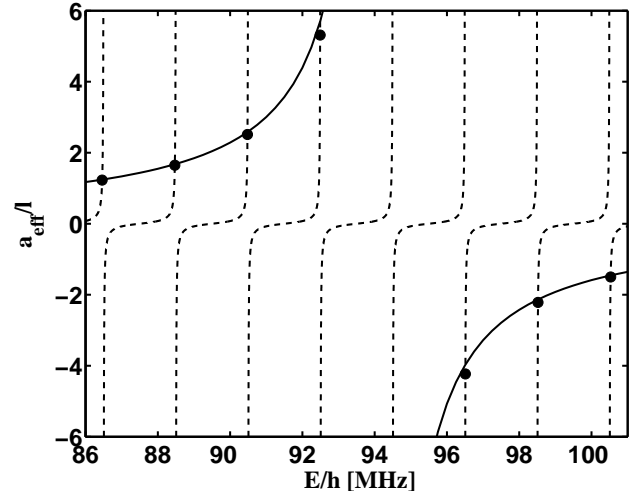} }
     \caption{Same as Fig.~\ref{selfconsplot1} but at a higher energy
     range.  Note that the effective scattering length diverges, but 
the
     self-consistent eigenvalues still agree with the exact numerical
     eigenvalues. }
 \label{selfconsplot2}
 \end{figure}

The first problem we consider is that of doubly polarized (electron
and nuclear spin up) $^{23}$Na atoms colliding in the trap.  In this 
case, there is only one scattering channel, governed by the $a^3 
\Sigma_u^{+}$ adiabatic Born-Oppenheimer potential.  The scattering 
length is $a = 3.2$ nm, and Figs.\  \ref{selfconsplot1} and 
\ref{selfconsplot2} show the effective scattering length as a 
function of energy.  It increases with energy and diverges near $E/h 
= 90$ MHz where $\delta_0 = \pi/2$ (this corresponds to a local 
maximum of the $s$-wave cross section), and is negative immediately 
above this energy.  In this work both the single- and multi-channel 
phase shifts are calculated by applying the Gordon propagation method 
\cite{Gordonmethod} with the best available  scattering potentials 
for Na$_2$~\cite{Samuelis01}.

The radial Schr{\"o}dinger equation for the Hamiltonian 
Eq.~(\ref{Hamiltonian}) was solved numerically for the eigenvalues.  
For 
a detailed description of our numerical method using a discrete 
variable representation, see Ref. \cite{DVRmethod}.
We take a trap frequency  of $\omega/ 2 \pi = 1$ MHz, for which 
$l=29.6$ nm and $\hbar \omega/k_B =48$ $\mu$K ($k_B$ is the Boltzmann 
constant).  Such a tight trap should be feasible in a Na optical 
lattice.

We illustrate the graphical solution of the self-consistent model in
Figs.  \ref{selfconsplot1} and \ref{selfconsplot2}.  In each plot, the
solid curve represents the left hand side of Eq.~(\ref{selfconsEq}),
$a_{\mbox{eff}}/l$, while the dashed curve is the right hand side. 
The abscissae of the points where the curves intersect give the
self-consistent eigenvalues according to the model.  One way of
comparing with the exact numerical eigenvalues is to evaluate the
intercept function $f(E)$ at these eigenvalues; these points are
plotted as circles.  The closer the circles lie to the intersection of
the curves, the better the agreement.  The exact numerical and model
eigenvalues in Fig.  \ref{selfconsplot1} agree to better than $0.0016
\, \hbar \omega$.  Note that the solution of 
Eq.~(\ref{analyticeigen}) is found from the intersection of $f(E)$ 
and the
horizontal line $E = a/l$; the corresponding eigenvalues differ
significantly from both the exact and self-consistent ones.

The range of energies in Fig.\ \ref{selfconsplot2} is centered near
the energy at which $a_{\mbox{eff}}/l$ diverges.  Even though
$|a_{\mbox{eff}}| \gg l$, the self-consistent eigenvalues are still
accurate.  They agree with the exact ones to within $0.0018 \, \hbar
\omega$.  Clearly this validates our model.  We have also obtained
eigenvalues for much higher trap frequencies, at which distortion of
the collision potential is expected to cause the self-consistent model
to fail.  At a trapping frequency $100$ MHz, where $l = 2.96$ 
nm $\approx x_0$ , the error between the exact eigenvalues and
those obtained from our model has increased to $0.045 \, \hbar
\omega$.  The crucial interaction length scale for comparison to the
trap size $l$ is $x_0$, not the effective scattering length $a_{\mbox
{eff}}$.

\begin{figure}[tb]
\noindent\centerline{ 
\epsfig{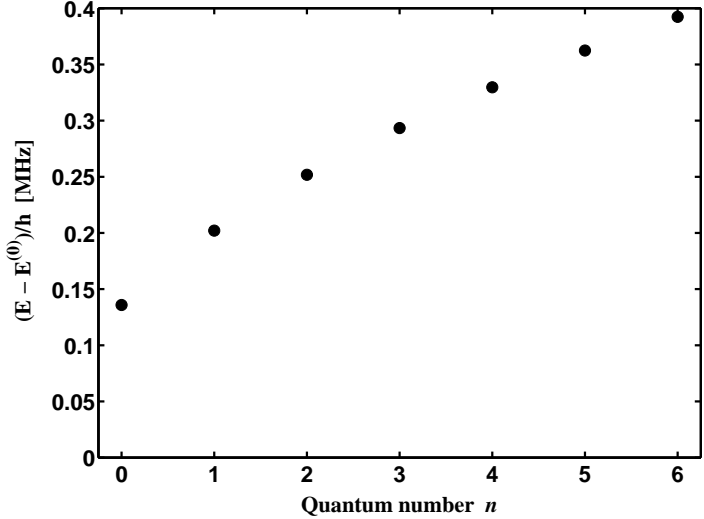} }
    \caption{Difference between eigenvalues for interacting and
    noninteracting, doubly polarized Na atoms in a 1 MHz trap versus
    quantum number $n$.  The self-consistent and exact numerical
    eigenvalues are indistinguishable on the scale of this figure. }
\label{eigvsindex_single}
\end{figure}

The difference between the lowest seven eigenvalues and the
corresponding harmonic oscillator eigenvalues given by 
Eq.~(\ref{unperturbE}) is plotted in Fig.  \ref{eigvsindex_single} 
versus
the quantum number $n$.  The shift due to the interactions is a
significant fraction of $\hbar \omega$ and should be observable in
appropriate experiments.  The dependence of the shift on the index for
the lowest few eigenvalues is due mainly to the energy dependence of
the gamma functions in Eq.~(\ref{interceptfn}), and only slightly due
to the variation of the effective scattering length with energy.  On 
the
other hand, for the higher eigenvalues in Fig.  \ref{selfconsplot2},
the shifts in eigenvalues arise mostly from the rapid variation of
effective scattering length with energy.  Near the asymptote
$a_{\mbox{eff}} \rightarrow \infty$ the eigenvalues have increased by
approximately $\hbar \omega$ compared with the unperturbed values.

The above examples show that accurate eigenvalues can be obtained by
using results of the single-channel scattering problem (without the
trap), and solving the self-consistent Eq.~(\ref{selfconsEq}).  Our
self-consistent model is good even when the effective scattering
length is large compared to the trap width, provided the trap size is 
still larger than the van der Waals length scale.

\section{multi-channel scattering and Feshbach resonance}
\label{multichannelsect}

In the previous Section, large ratios of effective scattering length
to trap width were only possible for very high-lying levels. 
Here we want to discuss a  situation where $|a_{\mbox{eff}}|/l$ 
is arbitrarily large for the lowest  trap levels.  This can be 
experimentally realized for $s$-wave collisions using a 
magnetically-tuned Feshbach resonance.

We  consider a  Feshbach resonance in the collision of two $^{23}$Na 
atoms 
in their lowest hyperfine level at a magnetic field near 90 mT 
\cite{Inouye98,vanAbeelen99,Mies00}.  The hyperfine states of the 
$^{23}$Na
atom diagonalize the Zeeman and hyperfine interaction and are labeled 
by 
$\ket{a}, \ket{b}, \ldots \ket{h}$, starting from the lowest internal
energy.  For very low collision energy, $s$-wave collisions of two 
$\ket{a}$ atoms are represented by five symmetrized asymptotic 
collision channels, one of which is open, $\ket{\{aa\}}$, and four of 
which are closed, $\ket{\{ag\}}$, $\ket{\{bh\}}$, $\ket{\{fh\}}$, and 
$\ket{\{gg\}}$.  The interaction between the
atoms is mediated by the X$^1\Sigma_g^+$ and a$^3\Sigma_u^+$
adiabatic Born-Oppenheimer potentials.  During the collision this
interaction mixes hyperfine states and is described by a
Hamiltonian coupling the above five channels \cite{Mies00}.  A 
Feshbach resonance 
state at energy $E_F$ is located at the
threshold of the $\ket{\{aa\}}$ channel for a magnetic field
$B_{\mbox{res}} \approx 90.09$ mT.  This resonance is a quasibound 
molecular  eigenstate of the four closed channel problem.   It can be 
formed from or decay to the $\ket{\{aa\}}$ open
channel, to which it is coupled.  As the magnetic field 
$B$ is changed near $B_{\mbox{res}}$, the
resonance energy also varies with $B$:
\be
E_F =  \fd{E_F}{B}  (B - B_{\mbox{res}}) .
\label{quasiboundB}
\ee

\begin{figure}[tb]
\noindent\centerline{ 
\epsfig{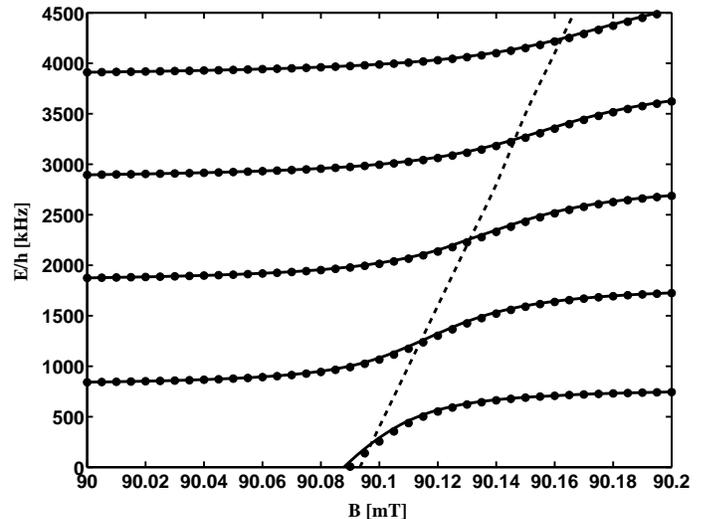} }
    \caption{Numerical {\em (circles)} and self-consistent {\em (solid
    curve)} eigenvalues vs magnetic field $B$ for Na in a 500 kHz 
   trap.  The dashed line shows $E_{\mbox{div}}$, where the effective 
   scattering length diverges  at a fixed value of magnetic field. }
\label{eigvsB}
\end{figure}

The analytic theory of Feshbach resonances  
\cite{Timmermans99a,Moerdijk95} shows that the phase shift $\delta_0$ 
can be written as the sum of background and resonant scattering 
contributions:
\be
  \delta_0 = \delta_{\mbox{bg}} - \arctan{\frac{\Gamma_F}{2(E - E_F - 
\Delta_F )}} ,
\ee
where $\Gamma_F$ is the linewidth, $\Delta_F$ is a level shift induced
by the coupling between the open and closed channels, and
$\delta_{\mbox{bg}}$ is the background phase shift.  It follows that 
the effective scattering 
length Eq.~(\ref{aeff}) for the $\{aa\}$ channel is 
\be
a_{\mbox {eff}}(E) = \frac{\frac{\Gamma_F}{2} - (E - E_F - \Delta_F ) 
\tan \delta_{\mbox{bg}} }
    {k \left( 
         E - E_F - \Delta_F + \frac{\Gamma_F}{2} \tan 
\delta_{\mbox{bg}}
     \right)} .
     \label{aeffFB}
\ee
  Up to the highest
energy we will consider, $E/h\approx$ 5 MHz, both $\Gamma_F$ and
$\tan \delta_{\mbox{bg}}$ are proportional to $\sqrt{E}$, and 
$\Delta_F$
becomes constant.   Moreover, Eq.~(\ref{aeffFB}) shows that the
effective scattering length diverges near the energy
\be E_{\mbox{div}} = 
\frac{E_F + \Delta_F}
     {1 + \frac{1}{2} \fd{}{E}
       \left( \Gamma_F \tan \delta_{\mbox{bg}} \right)_{E \rightarrow 
0}}, 
\ee 
The effective scattering length is positive below and negative above
$E_{\mbox{div}}$, which is magnetically tunable according to
Eq.~(\ref{quasiboundB}).  However, instead of employing the analytic
theory, at a given value of magnetic field we directly obtain 
the
effective scattering length as a function of $E$ from a numerical 
close
coupled scattering calculation with five channels.  This enables us to
extract the position of the divergence $E_{\mbox{div}}$, which is 
plotted 
as the dashed curve in the $(E, B)$ plane in Fig.~\ref{eigvsB}.

We now examine the effect of the Feshbach resonance on trap 
eigenstates, assuming a trap with $\omega/2\pi = 500$ kHz.  We used 
the numerical discrete variable method for five channels to calculate 
the lowest eigenvalues of the trap states for a range of magnetic 
fields near the
resonance.  These eigenvalues are plotted as circles in Fig.  
\ref{eigvsB}. 
Solutions to the self-consistent eigenvalues were obtained by solving
Eq.~(\ref{selfconsEq}) graphically as in the single-channel case; the
solutions versus magnetic field are the solid curves.  The 
self-consistent
eigenvalues agree well with the numerical ones for all values of 
energy and magnetic field; the worst agreement, $< 0.1\hbar 
\omega$, is for eigenvalues near the resonance position.   The 
self-consistent eigenvalues
always lie slightly above the numerical ones.  Note that the 
self-consistent
eigenvalues cross the $E_{\mbox{div}}$ curve near  $E/\hbar \omega =
1/2, 5/2, 9/2 \ldots$  Another particular feature of the plot, which 
is correctly
reproduced by the numerical solution, is that as $B$ decreases the 
lowest trap state ($E > 0$) becomes the highest  bound state ($E<0$)
 for a magnetic field $B < B_{\mbox{res}}$.  This occurs where the
effective scattering length is still finite and positive, since 
$a_{\mbox{eff}}/l \approx 1.48$ when $E=0$ in 
Eqs.~(\ref{interceptfn})-(\ref{selfconsEq}).

\section{conclusion} 
\label{conclusion}

We have shown how a self-consistent model can be used to calculate the
eigenvalues of interacting atoms in an isotropic harmonic trap.  Our
model involves solving an equation containing the effective scattering
length for untrapped atoms, and the trap frequency.  We compared our
model with exact results for $^{23}$Na both for a single channel 
collision
and a multi-channel collision with a tunable Feshbach resonance.  In 
both cases, the model can accurately treat tight traps, as long as 
the trap size is larger than the van der Waals scale length.  
Consequently we expect the model to apply to other atomic species.  
In particular,  Cs would be an interesting case for which the 
scattering length is large in comparison with even modest trap sizes 
\cite{Leo00}.

In the future, we want to generalize the self-consistent model to more
arbitrary trap potentials.  There are two technical problems to be
overcome.  First, for atomic collisions in anisotropic harmonic traps,
the relative coordinate equation does not separate; this implies that 
different partial waves are coupled via the anisotropy.   A related 
point is that the scattering of higher partial waves can
also be modeled by pseudopotentials \cite{Huang57}.  Second, for 
anharmonic traps, the center-of-mass and relative
atomic coordinates do not separate, and even more coordinates must be 
treated simultaneously.  Anharmonic terms become important for low 
intensity optical lattices or for trap levels with many quanta of 
excitation.

One would expect to be able to use the effective scattering length in 
many-body problems, where the pseudopotential approximation has had 
widespread use.  One would simply need to replace $a$ by the 
effective scattering length.  This should be especially useful and 
necessary for situations where a tunable Feshbach resonance is used 
to alter the interaction properties.  There are a number of cases 
where the relative collision energy  for a many-body system is 
well-defined, such as for condensates in optical lattices 
\cite{Greiner02}, colliding condensates \cite{Band00,Kohler01}, or 
cold gases of mixed fermionic species, where collisions occur at the 
Fermi energy.  It should also be possible to incorporate inelastic 
collision loss channels by using a complex effective scattering 
length \cite{Band00}.

\section*{acknowledgments}
ELB was supported by a fellowship from the National Research Council. 
ET and PSJ acknowledge support from the Office of Naval Research. 
Discussions with C. Greene, F. Mies, C. Williams, and B. Gao helped 
stimulate the work.

\bibliographystyle{prsty}
\bibliography{erics}

\end{multicols}
\end{document}